\documentclass[proof]{WileyASNA-v1}

\articletype{Article Type}%

\def\apjl{ApJ }%
\def\apjs{ApJS }%
\def\ao{Appl.~Opt.}%
\def\aap{A\&A }%
\received{15 September 2018}
\begin{document}

\title{$\eta$ Carinae: particle acceleration and multi-messenger aspects}

\author[1]{Roland WALTER}
\author[2]{Matteo BALBO}

\authormark{Walter, R., Balbo, M.}

\address{\orgdiv{Department of Astronomy}\\ \orgname{University of Geneva}\\ \orgaddress{Chemin d'Ecogia 16\\1290 Versoix\\ \country{Switzerland}\\ \email{roland.walter@unige.ch}}}

\abstract{$\eta$ Carinae is composed of two very massive stars orbiting each other in 5.5 years. The primary star features the densest known stellar wind, colliding with that expelled by its companion. The wind collision region dissipates energy and accelerate particles up to relativistic energies, producing non thermal X- and $\gamma$-ray emission detected by Beppo-SAX, INTEGRAL, Swift, Suzaku, Agile, Fermi and H.E.S.S.. The orbital variability of the system provides key diagnostic on the physics involved and on the emission mechanisms. The low-energy component, which cuts off below 10 GeV and varies by a factor $< 2$ along the orbit, is likely of inverse Compton origin. The high energy component varies by larger factors and differently during the two periastrons observed by Fermi. These variations match the predictions of simulations assuming a magnetic field in the range 0.4-1 kG at the surface of the primary star. The high-energy component and the thermal X-ray emission were weaker than expected around the 2014 periastron suggesting a modification of the inner wind density. Diffuse shock acceleration in the complex geometry of the wind collision zone provides a convincing match to the observations and new diagnostic tools to probe the geometry and energetics of the system. A future instrument sensitive in the MeV energy range could discriminate between lepto-hadronic and hadronic models for the gamma-ray emission. At higher energies, the Cherenkov Telescope Array will distinguish orbital modulations of the high-energy component from these of ultraviolet-TeV photo absorption providing a wealth of information constraining acceleration physics in more extreme conditions than found in SNR.}

\maketitle

\section{Introduction}

$\eta$ Carinae is the most luminous massive binary system of our galaxy and the first one to have been detected at very high energies, without hosting a compact object. It is composed by one among the most massive stars known ($\eta$ Car A) with an initial mass estimated above M$_{\rm{A}} \gtrsim 90 M_{\odot}$ \citep{2001ApJ...553..837H} and a companion ($\eta$ Car B) believed to be an O supergiant or a WR star. 
$\eta$ Car A is accelerating a very dense wind with a mass loss rate of $\sim 8.5\cdot 10^{-4}$ M$_{\odot}$ yr$^{-1}$ and a terminal wind velocity of $\sim 420$ km s$^{-1}$ \citep{2012MNRAS.423.1623G}. Its companion probably emits a fast low-density wind at $10^{-5}$ M$_{\odot}$ yr$^{-1}$ reaching a velocity of 3000 km s$^{-1}$ \citep{2002A&A...383..636P,2005ApJ...624..973V,2009MNRAS.394.1758P}. 

During its Great Eruption (1837-1856), $\eta$ Carinae experienced a huge outburst ejecting an impressive quantity of mass estimated as $10-40~M_\odot$ \citep{2010MNRAS.401L..48G} at an average speed of $\sim 650$ km s$^{-1}$ \citep{2003AJ....125.1458S} subsequently forming the Homunculus Nebula, and became one of the brightest stars of the sky. The regular modulation detected in the X-ray lightcurve suggests that the two stars are located in a very eccentric orbit \citep{2001ApJ...547.1034C,2008MNRAS.388L..39O}. The estimated orbital period at the epoch of the Great Eruption was $\sim$~5.1 yr, and increased up to the current $\sim~5.54$~yr \citep{2004MNRAS.352..447W,2005AJ....129.2018C,2008MNRAS.384.1649D}. 

Given the high eccentricity of the orbit, the relative separation of the two stars varies by a factor $\sim20$, reaching its minimum at periastron, when the two objects pass within a few AU of each other (the radius of the primary star is estimated as 0.5 AU). In these extreme conditions their supersonic winds interact forming a colliding wind region of hot shocked gas where charged particles can be accelerated up to high energies via diffusive shock acceleration \citep{1993ApJ...402..271E,2003A&A...409..217D,2006ApJ...644.1118R}. As these particles encounter conditions that vary with the orbital phase of the binary system, one can expect a similar dependency of their non thermal emission.

The hard X-ray emission detected by INTEGRAL \citep{2008A&A...477L..29L} and Suzaku \citep{2008MNRAS.388L..39O}, with an average luminosity $(4$-$7)\times10^{33}$ erg s$^{-1}$, suggested the presence of relativistic particles in the system. The following year AGILE detected a variable $\gamma$-ray source at the position of $\eta$ Carinae \citep{2009ApJ...698L.142T}. Other $\gamma$-ray analyses followed, reporting a luminosity of $1.6\times10^{35}$ erg s$^{-1}$ \citep{2010ApJ...723..649A,2011A&A...526A..57F,2012A&A...544A..98R}, and suggested the presence of a hard component in the spectrum around periastron, which subsequently disappeared around apastron. Such a component has been explained through $\pi^0$-decay of accelerated hadrons interacting with the dense stellar wind \citep{2011A&A...526A..57F}, or interpreted as a consequence of $\gamma$-ray absorption against an ad hoc distribution of soft X-ray photons \citep{2012A&A...544A..98R}.

\begin{figure}[]
\includegraphics[width=0.49\textwidth]{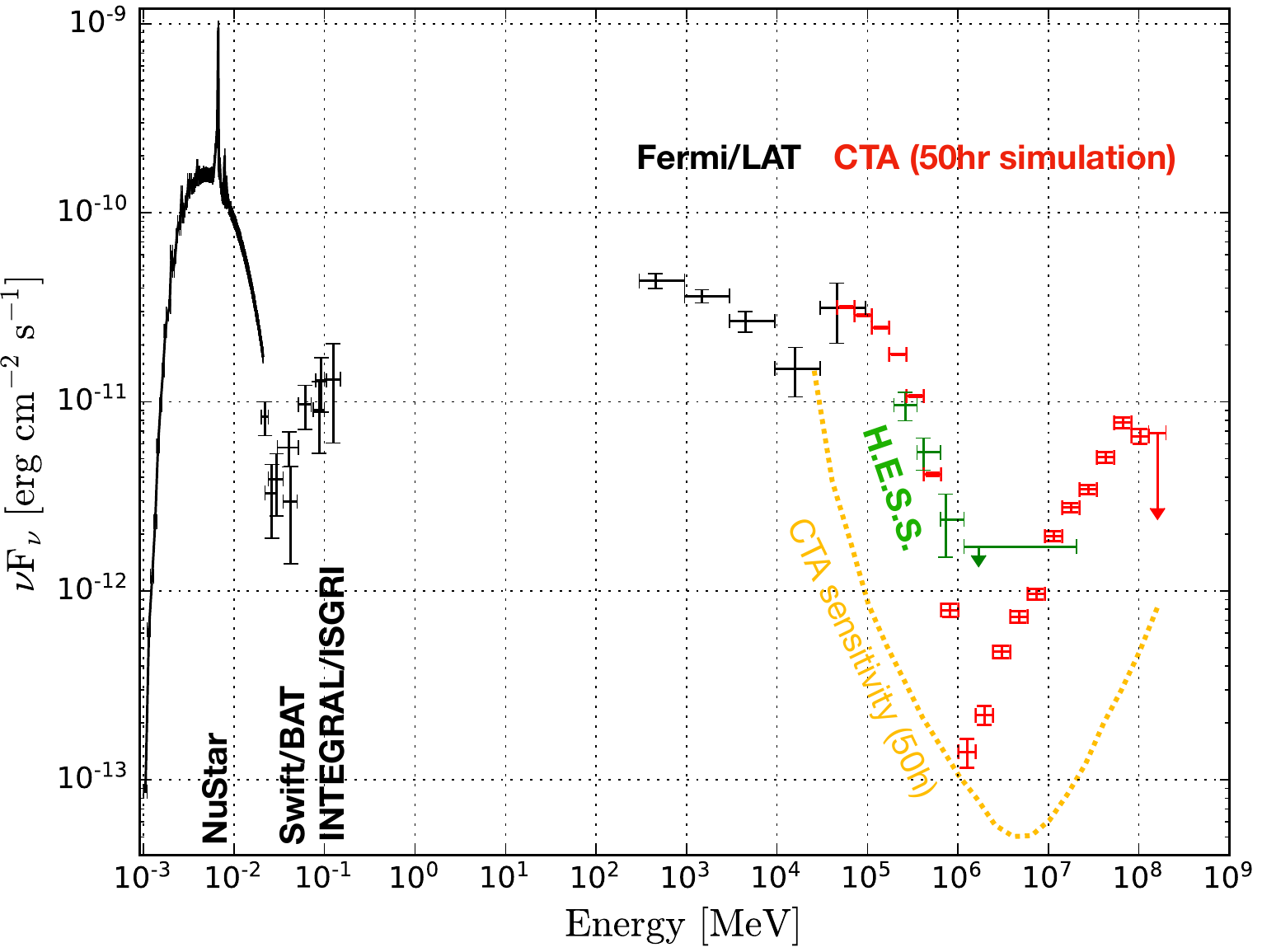}
\caption{Spectral energy distribution of $\eta$ Carinae from 1 keV to 100 TeV. The data are from NuStar \citep{2018A&A...610A..37P}, Swift/BAT, INTEGRAL, Fermi/LAT and H.E.S.S. and obtained close to periastron. The red points show the results of a simulation of what could be detected by CTA (at periastron) assuming that the emission is dominated by $\pi_0$ decay modified by photo-photo absorption in the strong ultraviolet photon field.}\label{fig:sed2}
\end{figure}

H.E.S.S. detected $\eta$ Carinae  \citep{2017arXiv170801033L} providing additional constrains on its spectral energy distribution (Fig. \ref{fig:sed2}). The emission, from the wind shock region, is dominated by thermal emission in the soft X-rays, inverse Compton emission from 100 keV to 10 GeV and by an additional component up to TeVs. 

\section{Orbital variability in the GeV band}

\cite{2011ApJ...726..105P} presented three dimensional hydrodynamical simulations of $\eta$ Carinae including radiative driving of the stellar winds \citep{1975ApJ...195..157C}, optically-thin radiative cooling \citep{2000adnx.conf..161K}, gravity and orbital motion. The main aim of these simulations was to reproduce the X-ray emission analysing the emissivity and the self-obscuration of the stellar wind. The simulations reproduced the observed X-ray spectra and lightcurves reasonably well, excepting the post-periastron extended X-ray minimum, where the flux was overestimated. Additional gas cooling, e.g. by particle acceleration and inverse-Compton processes, could decrease the wind speed and increase the cooling and disruption of the central wind collision zone. 

To estimate the non thermal emission predicted by these simulations \citep{2017A&A...603A.111B}  calculated the maximum energies that could be reached by electrons and hadrons \citep{2011A&A...526A..57F} cell-by-cell assuming a dipolar magnetic field at the surface of the main star, perpendicular to the orbital plane. The magnetic field is the only additional parameter with respect to the simulations by \cite{2011ApJ...726..105P}. Shock velocities and mechanical power were calculated in every cell, including those outside the shock region. As expected, most of the shock power is released on both sides of the wind collision zone and in the cells downstream the wind-collision region \citep{2006ApJ...644.1118R}. The evidence that the X-ray luminosity at apastron is about a third of the peak emission at periastron can be explained since the increasing shock area compensates the loss of the released energy density up to a relatively large distance from the center of the wind collision zone.

The mechanical luminosity available in the shock increases towards periastron (the same trend is followed by the X-ray thermal emission) and almost doubles in the phase range $\sim 1.05 - 1.15$. The latter peak corresponds to a bubble with reverse wind conditions developing because of the orbital motion, effectively doubling the shock front area during about a tenth of the orbit \citep{2011ApJ...726..105P}. The density of this bubble is much lower than that of the central wind collision zone and does not contribute much to the thermal X-ray emission. The mechanical luminosity shows a local minimum between phases 1.0 and 1.05, when the central part of the wind collision zone is disrupted. 

\begin{figure}[]
\includegraphics[width=0.49\textwidth]{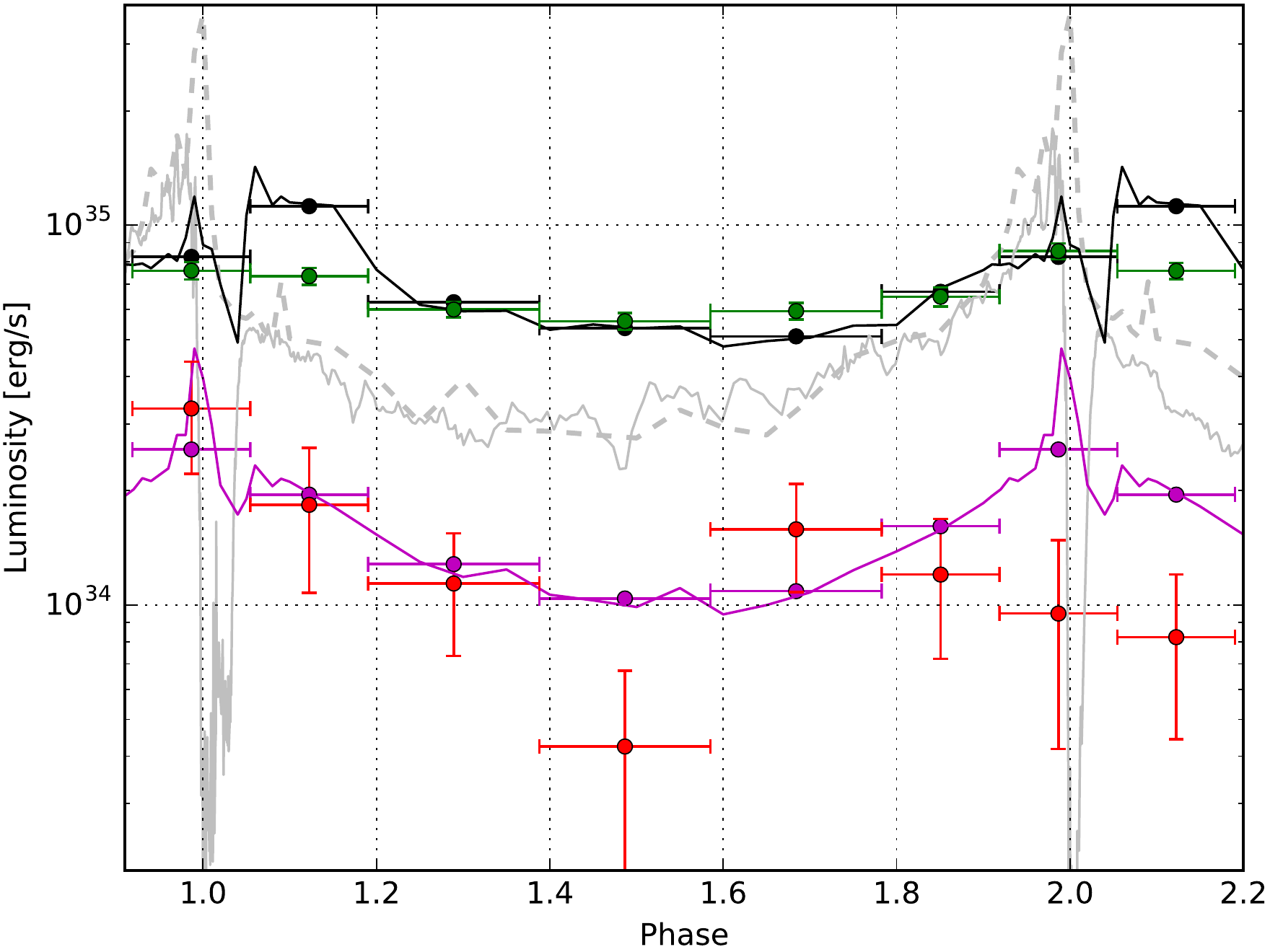}
\caption{Simulated and observed X-ray and $\gamma$-ray lightcurves of $\eta$ Carinae. The black and purple lines and bins show the predicted inverse-Compton and neutral pion decay lightcurves. The green and red points show the observed Fermi-LAT lightcurves at low (0.3-10 GeV) and high (10-300 GeV) energies. The dim grey lightcurves show the observed (continuous) and predicted (dash, without obscuration) thermal X-ray lightcurves. Error bars are $1\sigma$.}\label{fig:simul}
\end{figure}

Electron cooling, through inverse-Compton scattering, is very efficient, thus $\gamma$-rays are expected to peak just before periastron. A secondary inverse-Compton peak could be expected above phase 1.05 with a different spectral shape compared with the former as the UV seed thermal photons will have lower density when compared to the location of the primary shock close to the center of the system. In our simplified model we assumed that the spectral shape of the seed photons is the same in all cells of the simulation (r$^{-2}$ dependency is taken into account), and that these soft photons are enough to cool down all relativistic electrons. The relative importance of the second peak, however, depends on the magnetic field geometry, radiation transfer (neglected in our model), obscuration and details of the hydrodynamics (which do not represent the soft X-ray observations very well in this phase range). These details are not well constrained by the available observations and we did not try to refine them. Figure \ref{fig:simul} shows the observed and predicted X and $\gamma$-ray lightcurves. To ease the comparison between observations and simulations, the results of the latter were binned in the same way as the observed data.

\begin{figure}[]
\includegraphics[width=0.49\textwidth]{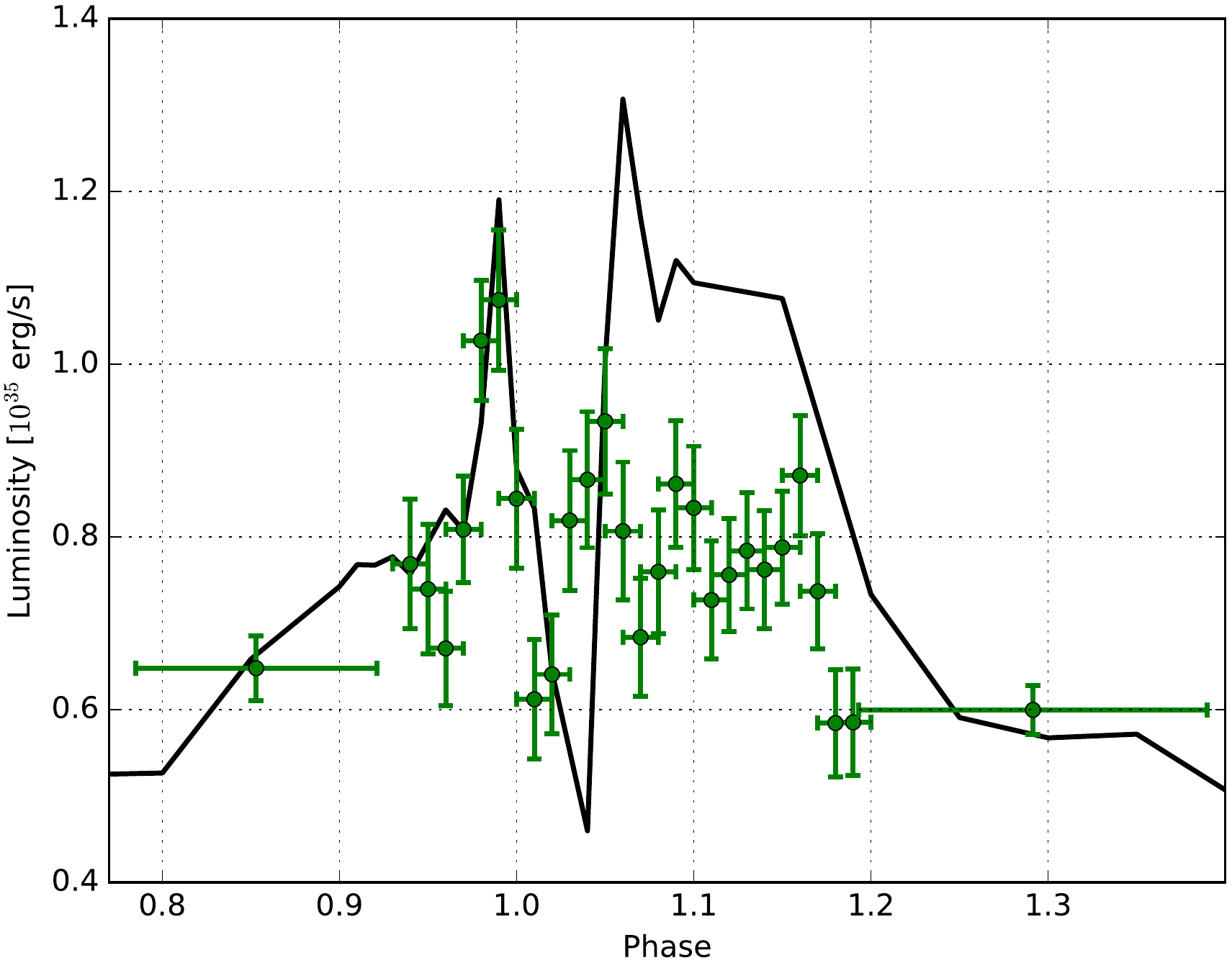}
\caption{A merged Fermi LAT analysis (0.3-10 GeV) of the two periastrons for narrow time bins. The two broad bins and the black curve are the same as in Fig. \ref{fig:simul}.}\label{fig:peri}
\end{figure}

Both the predicted inverse-Compton emission and the observed (0.3-10 GeV) LAT lightcurve show a broad peak extending on both sides of periastron, as expected from the evolving shock geometry. The amplitude of the variability in the simulation depend on the number/size of those cells where particles can be accelerated up to relevant energies, which in turn depends on the magnetic field. Probing the range suggested by \cite{2012SSRv..166..145W}, a surface magnetic field larger than 500~G provides a good match to the observations, while lower fields produce too large variations. We have not considered any magnetic field amplification at the shock, which in turn could obviously scale down the surface magnetic field required to get equivalent results. These comparisons indicates that about 1.5\% and 2.4\% of the mechanical energy is used to accelerate electrons and protons, respectively. Assuming a field of 500 G for the rest of the discussion, the predicted flux at phase 1.1 is twice too large when compared with the observation. This discrepancy largely comes from the energy released in the inverted wind bubble after periastron. The ratio of the emission generated in the shocks on both sides of the wind collision zone is relatively constant along the orbit excepting at phase 1.1, where much more power is generated in the shock occurring in the wind of the secondary star. The inverted bubble might either be unstable in reality or produce a significantly different inverse-Compton spectrum.

Since the low energy spectra during both periastrons are similar, we analysed simultaneously the Fermi LAT low energy data derived from the two periastrons, binned in shorter time intervals (Fig.~\ref{fig:peri}). They show a peak at periastron, a minimum at phase 1.02 and a second broad peak at phase 1.1, similar to the predictions of the simulations. The only differences are that the observed second broad peak is slightly shifted towards earlier phases and has a lower luminosity when compared to the simulations. The similarities between the observations and the simulations, $\gamma$-ray peak and minimum with consistent duration and amplitude, are very encouraging. The phase difference could be related to the eccentricity $(\epsilon=0.9)$ assumed in the simulations, which is not well constrained observationally \citep{2000ApJ...528L.101D,2001ApJ...547.1034C} and that has an important effect on the inner shock geometry. 

\begin{figure}[]
\includegraphics[width=0.49\textwidth]{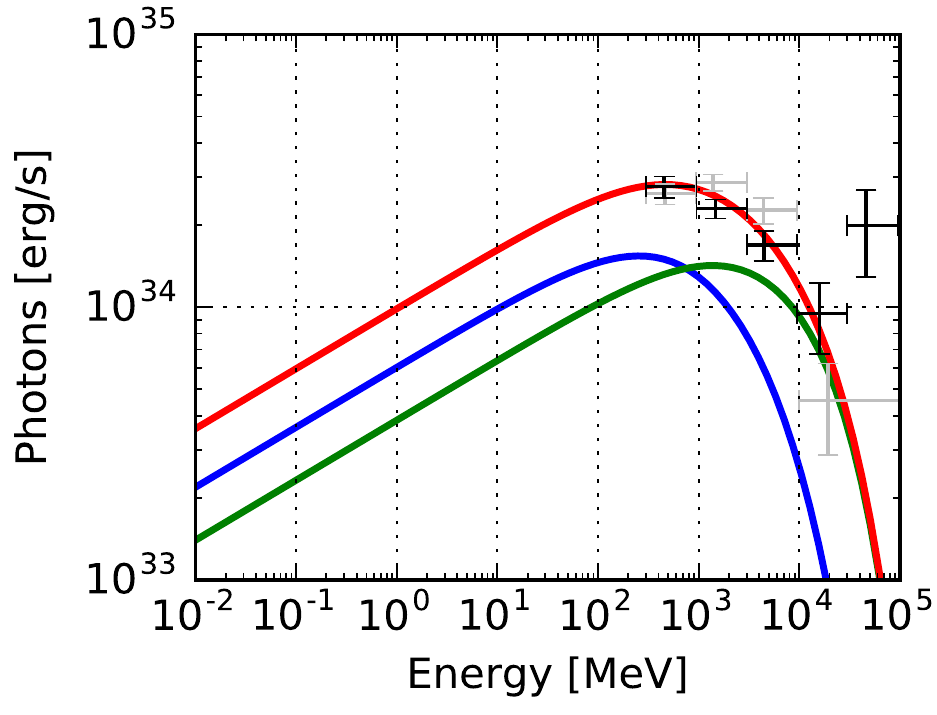}
\caption{Inverse Compton spectra produced by electrons accelerated in the shocks occurring on both sides of the wind separation surface (in green and blue). Their sum (in red) shows up as a single smoothed component. The data points are from Fermi LAT for the two observed periastrons.}\label{fig:espec}
\end{figure}

The distribution of the electron Lorentz factor $\gamma_e$, weighted by the emissivity, is relatively smooth and the expected photon distribution is very smooth. The difference of the electron spectral shape on both sides of the wind collision zone (Fig.: \ref{fig:espec}) cannot explain the two components $\gamma$-ray emission as suggested by \cite{2011A&A...530A..49B}, who assumed a simplified geometry. We obtain a good match between the observed low energy $\gamma$-ray spectrum and the predictions of the simulations at periastron, even though some discrepancy can be observed at apastron where an excess is observed between 2 and 10 GeV.

The inverse-Compton emission peaks slightly below 1 GeV and does not extend beyond 10 GeV at a level consistent with the observations during the first periastron, contrasting with the conclusions from \cite{2015MNRAS.449L.132O}, attributing the full Fermi LAT detection to hadronic emission. Their simulations predict a smaller variation between periastron and apastron, a longer flare around periastron and a deeper minimum when compared to the observed data. Such discrepancies might be due to the simplified geometry assumed by the authors and by the artificially reduced particle acceleration at periastron. Inverse-Compton emission and neutral pion decay \citep{2011A&A...526A..57F} remains therefore a very good candidate to explain the Fermi observations.

An instrument sensitive in the 1-100 MeV band, such as e-Astrogam \citep{2017ExA...tmp...24D}, will easily discriminate between the lepto-hadronic and the hadronic models for the gamma-ray emission, as the inverse Compton leptonic emission of the former would be much stronger than predicted by the latter, and therefore strongly constrain the acceleration physics in more extreme conditions than found in SNR.

\section{Extrapolation at very high energies}

The situation is differen t for hadrons. Unless the magnetic field would be very strong ($>$ kG) hadronic interactions mostly take place close to the center and a single peak of neutral pion decay is expected before periastron. The high-energy flux of $\eta$ Carinae, reported in Fig.~\ref{fig:simul}, decreased after the first periastron passage of 2009 towards apastron and did not increase again toward the periastron of 2014. The simulated pion induced $\gamma$-ray lightcurve and its variability amplitude show a single peak of emission centered at periastron, in good agreement with the Fermi LAT observations of the first periastron. The results of the observations of the second periastron are different, with a lack of emission. It has been suggested that the change of the X-ray emission after that periastron \citep[a significant decrease can be observed in Fig. \ref{fig:simul}, see also][] {2015arXiv150707961C} was the signature of a change of the wind geometry, possibly because of cooling instabilities. A stronger disruption or clumpier wind after the second periastron could perhaps induce a decrease of the average wind density and explain that less hadronic interactions and less thermal emission took place, without affecting  much inverse-Compton emission.

According to our simulations hadrons could be accelerated up to $10^{15}$ eV around periastron and reach $10^{14}$ eV on average. 
The choice of a lower magnetic field reduces those energies at apastron to $\sim 6 \times 10^{12}$~eV and $\sim 2 \times 10^{12}$~eV, and at periastron to $\sim 5.6 \times 10^{14}$~eV and $\sim 1.9 \times 10^{14}$~eV, respectively for 300~G and 100~G. $\eta$ Carinae can therefore probably accelerate particles close to the knee of the cosmic-ray spectrum. The spectra and the maximum particle energy depend of course on several assumptions, in particular the magnetic field. 

The highest energy $\gamma$-rays will be photo-absorbed (Fig. \ref{fig:sed2}) and a strong orbital modulation could be expected in the TeV domain. Depending on the assumed soft energy photons distribution and the consequent $\gamma$-$\gamma$ absorption at very high energy, $\eta$ Carinae could be detected by the CTA southern array at more than $10\sigma$ in spectral bins of $\Delta {\rm E/E} = 20\%$ for exposures of 50 hours, enough to measure separately the variability along the orbit of the high energy continuum and of photo absorption \citep{Acharya20133}. 

$\gamma$-ray observations can probe the magnetic field and shock acceleration in details. The quality of the current data above 1 GeV does not yet provide enough information to test hydrodynamical models including detailed radiation transfer (inverse-Compton, pion emission, photo-absorption). The interplay between disruption and obscuration does not yet account for the X-ray minimum and orbit to orbit variability. More sensitive $\gamma$-ray observations will provide a wealth of information and allow to test the conditions and the physics of the shocks at a high level of details, making of $\eta$ Carinae a perfect laboratory to study particle acceleration in wind collisions. 

\section{Acceleration physics and multi-messenger connections}

In the above section we have presented a model where electrons and protons are accelerated in the wind collision region of $\eta$ Carinae \citep[as initially proposed by][] {1993ApJ...402..271E}. $\eta$ Carinae could yield $10^{48-49}$ erg of cosmic-rays, a number close to the expectation for an average supernova remnant \citep{2016APh....81....1B}. The fraction of the shock mechanical luminosity accelerating electrons appears to be slightly smaller than that accelerating protons, contrasting with the efficiencies derived from the latest particle-in-cell 
simulations \citep{2015PhRvL.114h5003P}, however involving lower magnetic fields, radiation and particle densities and favouring hadronic acceleration in the context of SNR. 

Purely hadronic acceleration has been proposed \citep{2015MNRAS.449L.132O} to explain the GeV spectrum of Eta Carinae. In that case the two spectral components are related to the different hadron interaction times observed on the two sides of the wind separation surface, largely because of the contrast in density and magnetic field. In our simulations this effect is smoothed by the many zones of the model, each characterized by different conditions. Even if the shock on the companion side does contribute more at high energies, the predicted pion decay spectrum does not feature two components.

Assuming that for each photon originated via hadronic processes we also have the production of one neutrino, we derive a neutrino flux above 10 TeV that might reach $10^{-9}$ GeV s$^{-1}$cm$^{-2}$ on average, which is of the order of the IceCube neutrino sensitivity for several years of observations \citep{2017ApJ...835..151A}. Stacking some months of IceCube data obtained around periastron, over several decades could allow the detection of one PeV neutrino, above the atmospheric background. 

GeV-keV photo-absorption can in principle affect the observed spectral shape and create a local minimum close to 10 GeV \citep{2012A&A...544A..98R}. As the relevant optical depth remains negligible, this is however unlikely to play any significant role \citep{2017A&A...603A.111B} and would, in addition, require an excessively large cutoff energy.

Because of the very strong stellar photon field, TeV-ultraviolet photo-absorption is expected to be strong. The energy of the peak absorption (around 1 TeV at periastron) and the optical depth will vary along the orbit. CTA will be sensitive enough to detect these variations and will provide very valuable information on the system geometry. As the black body stellar radiation is narrow, CTA could measure independently the variations of the obscuration and of the intrinsic $\gamma$-ray continuum, i.e. of the particle spectrum. 

Ultraviolet-TeV photo-absorption will create electrons and positrons pairs far from the shock region at an energy of 0.1-1 TeV, similar to that of the excesses observed locally by Pamela, Fermi, AMS-02 and DAMPE. $\eta$ Carinae cannot explain the galactic cosmic-rays or the local electron/positron excess alone. However it shows that hydrodynamical shocks in the vicinity of massive stars can accelerate particles to very high energies and that $\gamma$-rays interacting with stellar light could in principle generate electron-positron pairs.

\bibliography{references}

\end{document}